\documentclass[sigconf,nonacm]{acmart}

\usepackage{tabularx}
\usepackage{booktabs}
\usepackage{graphicx}
\usepackage{amsmath}
\usepackage{microtype}
\usepackage{xcolor}
\usepackage{listings}
\usepackage{enumitem}
\usepackage{multirow}
\usepackage{array}
\usepackage{tikz}
\usepackage{pgfplots}
\usetikzlibrary{positioning,arrows.meta,fit,backgrounds}
\pgfplotsset{compat=1.18}
\makeatletter
\def\@authorfont{\large}
\def\@affiliationfont{\small}
\def\@ACM@checkaffil{}
\makeatother
\newcommand{\blfootnote}[1]{%
  \begingroup
  \renewcommand\thefootnote{}%
  \footnotetext[0]{#1}%
  \endgroup
}

\begin{document}

\title{Context as Infrastructure: Knowledge Compilation for Retrieval in Agentic Systems with Harnesses}

\author{Zhichang Yu\textsuperscript{1}}
\email{zhichang.yu@infiniflow.ai}
\affiliation{\mbox{}}

\author{Baining Cao\textsuperscript{2}}
\email{71295902003@stu.ecnu.edu.cn}
\affiliation{\mbox{}}

\author{Sbaaoui Idriss\textsuperscript{1}}
\email{idriss.sbaaoui@infiniflow.ai}
\affiliation{\mbox{}}

\author{Qi Wang\textsuperscript{1}}
\email{qi.wang@infiniflow.ai}
\affiliation{\mbox{}}

\author{Gangqiang Xu\textsuperscript{1}}
\email{gangqiang.xu@infiniflow.ai}
\affiliation{\mbox{}}

\author{Ling Qin\textsuperscript{1}}
\email{ling.qin@infiniflow.ai}
\affiliation{\mbox{}}

\author{Xiaoling Jin\textsuperscript{1}}
\email{xiaoling.jin@infiniflow.ai}
\affiliation{\mbox{}}

\author{An Liu\textsuperscript{1}}
\email{an.liu@infiniflow.ai}
\affiliation{\mbox{}}

\author{Shaohao Cai\textsuperscript{1}}
\email{shaohao.cai@infiniflow.ai}
\affiliation{\mbox{}}

\author{Xiaoyang Bao\textsuperscript{1}}
\email{xiaoyang.bao@infiniflow.ai}
\affiliation{\mbox{}}

\author{Shengle Chen\textsuperscript{1}}
\email{shengle.chen@infiniflow.ai}
\affiliation{\mbox{}}

\author{Changchang Ding\textsuperscript{1}}
\email{changchang.ding@infiniflow.ai}
\affiliation{\mbox{}}

\author{Pengfei Zhu\textsuperscript{1}}
\email{pengfei.zhu@infiniflow.ai}
\affiliation{\mbox{}}

\author{Jingxiang Xu\textsuperscript{1}}
\email{jingxiang.xu@infiniflow.ai}
\affiliation{\mbox{}}

\author{Yifeng Mao\textsuperscript{1}}
\email{yifeng.mao@infiniflow.ai}
\affiliation{\mbox{}}

\author{Lu Liu\textsuperscript{1}}
\email{lu.liu@infiniflow.ai}
\affiliation{\mbox{}}

\author{Hai Jin\textsuperscript{1}}
\email{hai.jin@infiniflow.ai}
\affiliation{\mbox{}}

\author{Yingfeng Zhang\textsuperscript{1}}
\email{yingfeng.zhang@infiniflow.ai}
\affiliation{\mbox{}}

\begin{abstract}
Retrieval-augmented generation is becoming agentic. A search harness writes queries, reads what it finds, reformulates, and decides when to stop, and this shift has largely solved the accuracy problem: a harness whose only retrieval tools are BM25 and grep, with MiniMax M3 as the answering model, answers 85\% of BrowseComp-Plus and reaches 91\% on the multi-hop FRAMES benchmark---two benchmarks whose figures are not comparable---in our runs, without any compiled index and without any vector index. What the shift exposes is a context problem. An undisciplined loop repeats the same queries, rereads the same documents, and fills the window with passages that are never used; the process consumes tokens but does not become an answer. We argue that the missing layer is \emph{knowledge compilation}: a configurable ingestion pipeline in which LLMs enrich source documents with context that no chunk contains on its own---navigation hierarchies whose leaves are atomic claims with verbatim evidence, entity graphs, timelines, mind maps, ontologies, wikis, and memory. Compilation is not a better retriever, and it does not need to be: under a competent harness, structural artifacts gave no detectable accuracy gain on FRAMES, and a PageIndex-style outline degraded at FRAMES scale. Compilation is a supply layer. A navigation tree whose leaves are claim--evidence pairs---the mechanism whose effect we can measure here---cuts total query tokens by 91\% when the evidence suffices and adds 1.9\% when it does not, for a 38\% average reduction; the asymmetry implies a 2\% break-even sufficiency rate and a measured rate near 43\%. A decision gate cuts latency from 140\,s to 30\,s, although the token saving of substitution is bought with retrieval rounds and carries a latency cost of its own. Ontology, graph, wiki, timeline, mind-map, and memory artifacts pay off in deployment---governance, cross-source integration, onboarding, temporal orientation, continuity---where accuracy benchmarks cannot see them. We describe RAGFlow's compilation pipeline and retrieval harness, which turns the loop into five disciplined stations and keeps the context entering the model the smallest sufficient set; the two together form what we call an agent-native database: a store that hands the agent compiled, governed context instead of a ranked list of passages.
\end{abstract}

\maketitle
\footnotetext[1]{InfiniFlow Inc., Shanghai, China}
\footnotetext[2]{East China Normal University, Shanghai, China}

\blfootnote{RAGFlow project: the system described in this paper is implemented in the open-source RAGFlow project, \url{https://github.com/infiniflow/ragflow}.}

\fancyhead[RE]{\sffamily\footnotesize Context as Infrastructure}
\fancyhead[LO]{\sffamily\footnotesize Context as Infrastructure}

\ccsdesc[500]{Information systems~Information retrieval}
\ccsdesc[300]{Information systems~Retrieval models and ranking}
\ccsdesc[300]{Computing methodologies~Natural language processing}

\keywords{Knowledge Compilation, Agentic Retrieval, Retrieval Harness, Context Engineering, Claim--Evidence Pairs, Ontology, Agent-Native Database, RAG}

\section{Introduction}
\label{sec:intro}

Retrieval-augmented generation (RAG) was introduced to ground language models in external knowledge~\citep{lewis2020rag}, but its center of gravity has moved. In a modern agentic system, retrieval is no longer a single query-to-embedding-to-similarity call. It is a loop in which an LLM writes lexical queries, reads snippets, reformulates, and decides when the collected evidence suffices~\citep{yao2023react,vijayaraghavan2025agenticrag,li2026beyond}. We call the component that owns this loop the \emph{search harness}, and its rise has changed what a retrieval layer is for: once the harness can grep a corpus and read what it finds, the classical objectives of the retrieval layer---better chunking, better embeddings, higher Recall@$k$---stop being the binding constraint. We build accordingly. This paper contains no vector index and no dense-retrieval leg, and the reason is empirical rather than aesthetic: adding a dense leg lowered accuracy in our runs, in the same direction as the reports of the teams cited below. A harness already performs the semantic step---the agent rewrites the query, reads what it finds, and states what is still missing---so a dense leg adds no signal the loop lacks; it adds a second ranking that competes with the lexical anchor the agent already trusts, at the price of a second index to build and refresh. BM25 and grep are what a harness needs, and every measurement here is taken with them. In our own runs a harness with only BM25, grep, and MiniMax M3 answers 85\% of BrowseComp-Plus correctly and reaches 91\% on the multi-hop FRAMES benchmark---two benchmarks whose figures are not comparable---without any compiled index, and others report the same saturation~\citep{subramanian2025grep,hsu2026piserini}.

The first claim of this paper is easy to state and hard to internalize: \textbf{retrieval is a harness problem} before it is an index problem. The index determines what can be found; the harness determines whether finding becomes an answer. A better index cannot fix a loop that never discards a bad query, reads without a budget, or keeps a passage because it was expensive to fetch.

The harness solves accuracy. It does not solve the \emph{process} of answering. An agentic loop is a search procedure with a language model inside it, and a procedure without discipline degenerates into blind exploration: it rewrites the same keywords, rereads the same document, piles dozens of marginal passages into the context, and finally bets on the model's patience for an answer. Every turn costs tokens, and every irrelevant passage competes with the relevant one for attention~\citep{liu2024lost}.

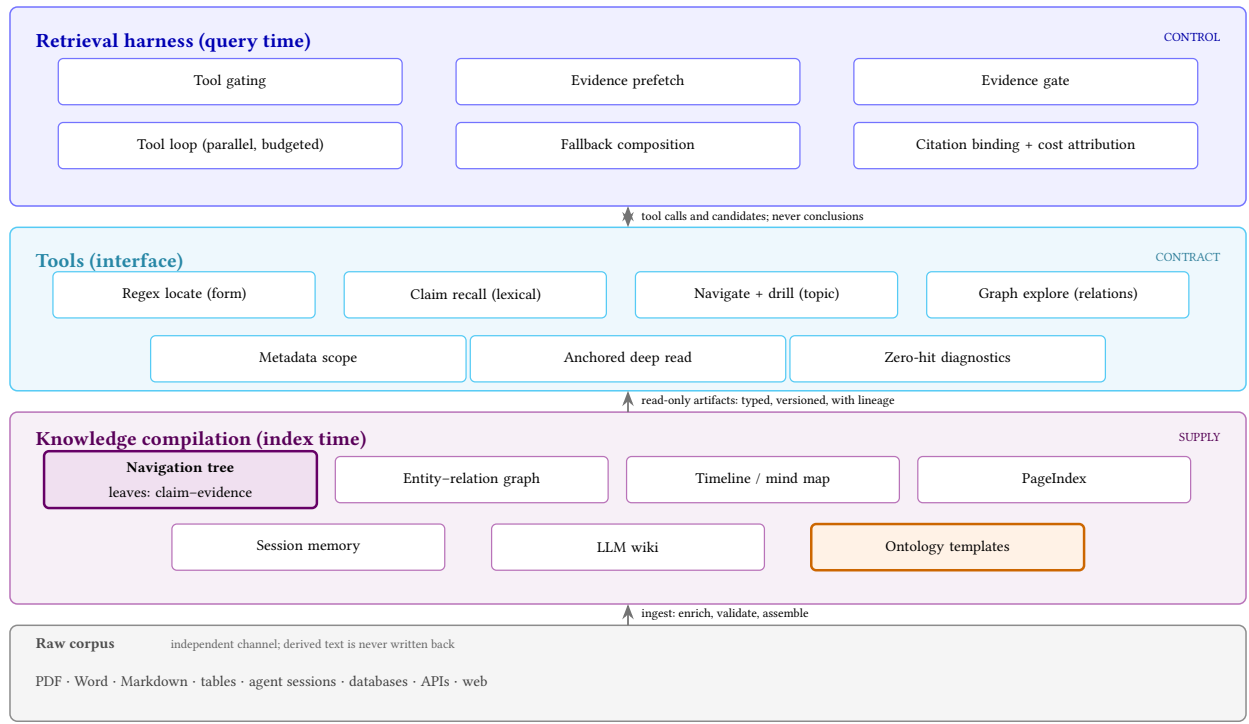
\begin{figure*}[t]
\centering
\resizebox{0.92\textwidth}{!}{%
\begin{tikzpicture}[font=\scriptsize,>=Stealth,
 band/.style 2 args={rounded corners=3pt,draw=#1,fill=#2,line width=0.6pt},
 chip/.style 2 args={rounded corners=2pt,draw=#1,fill=#2,align=center,inner sep=2pt,minimum height=6.5mm,line width=0.5pt},
 hot/.style={rounded corners=2pt,draw=violet!80!black,fill=violet!12,align=center,inner sep=2pt,minimum height=6.5mm,line width=1pt},
 onto/.style={rounded corners=2pt,draw=orange!80!black,fill=orange!10,align=center,inner sep=2pt,minimum height=6.5mm,line width=1pt},
 ar/.style={->,draw=black!55,line width=0.7pt},
 bi/.style={<->,draw=black!55,line width=0.7pt},
 flowlab/.style={font=\tiny,anchor=west,fill=white,inner sep=1pt}
]
\draw[band={blue!55}{blue!6}] (0,7.8) rectangle (17.4,10.6);
\node[anchor=north west,font=\small\bfseries,text=blue!70!black] at (0.25,10.35) {Retrieval harness (query time)};
\node[anchor=north east,font=\tiny,text=blue!65!black] at (17.15,10.35) {CONTROL};
\node[chip={blue!55}{white},text width=4.7cm] at (3.1,9.55) {Tool gating};
\node[chip={blue!55}{white},text width=4.7cm] at (8.7,9.55) {Evidence prefetch};
\node[chip={blue!55}{white},text width=4.7cm] at (14.3,9.55) {Evidence gate};
\node[chip={blue!55}{white},text width=4.7cm] at (3.1,8.65) {Tool loop (parallel, budgeted)};
\node[chip={blue!55}{white},text width=4.7cm] at (8.7,8.65) {Fallback composition};
\node[chip={blue!55}{white},text width=4.7cm] at (14.3,8.65) {Citation binding + cost attribution};
\draw[band={cyan!55}{cyan!6}] (0,5.2) rectangle (17.4,7.5);
\node[anchor=north west,font=\small\bfseries,text=cyan!60!black] at (0.25,7.25) {Tools (interface)};
\node[anchor=north east,font=\tiny,text=cyan!55!black] at (17.15,7.25) {CONTRACT};
\node[chip={cyan!55}{white},text width=3.55cm] at (2.45,6.55) {Regex locate (form)};
\node[chip={cyan!55}{white},text width=3.55cm] at (6.55,6.55) {Claim recall (lexical)};
\node[chip={cyan!55}{white},text width=3.55cm] at (10.65,6.55) {Navigate + drill (topic)};
\node[chip={cyan!55}{white},text width=3.55cm] at (14.75,6.55) {Graph explore (relations)};
\node[chip={cyan!55}{white},text width=4.3cm] at (4.2,5.65) {Metadata scope};
\node[chip={cyan!55}{white},text width=4.3cm] at (8.7,5.65) {Anchored deep read};
\node[chip={cyan!55}{white},text width=4.3cm] at (13.2,5.65) {Zero-hit diagnostics};
\draw[band={violet!55}{violet!6}] (0,2.2) rectangle (17.4,4.9);
\node[anchor=north west,font=\small\bfseries,text=violet!70!black] at (0.25,4.75) {Knowledge compilation (index time)};
\node[anchor=north east,font=\tiny,text=violet!65!black] at (17.15,4.72) {SUPPLY};
\node[hot,text width=3.7cm,minimum height=8.0mm] at (2.4,3.95) {\textbf{Navigation tree}\\[3pt] leaves: claim--evidence};
\node[chip={violet!55}{white},text width=3.7cm] at (6.5,3.95) {Entity--relation graph};
\node[chip={violet!55}{white},text width=3.7cm] at (10.6,3.95) {Timeline / mind map};
\node[chip={violet!55}{white},text width=3.7cm] at (14.7,3.95) {PageIndex};
\node[chip={violet!55}{white},text width=3.7cm] at (4.2,3.0) {Session memory};
\node[chip={violet!55}{white},text width=3.7cm] at (8.7,3.0) {LLM wiki};
\node[onto,text width=3.7cm] at (13.2,3.0) {Ontology templates};
\draw[band={black!45}{black!4}] (0,0.55) rectangle (17.4,1.9);
\node[anchor=west,font=\scriptsize\bfseries,text=black!70] at (0.25,1.62) {Raw corpus};
\node[anchor=west,font=\tiny,text=black!60] at (2.15,1.62) {independent channel; derived text is never written back};
\node[anchor=west,font=\scriptsize,text=black!75] at (0.25,1.1) {PDF $\cdot$ Word $\cdot$ Markdown $\cdot$ tables $\cdot$ agent sessions $\cdot$ databases $\cdot$ APIs $\cdot$ web};
\draw[ar] (8.7,1.9) -- (8.7,2.2);
\node[flowlab] at (8.85,2.05) {ingest: enrich, validate, assemble};
\draw[ar] (8.7,4.9) -- (8.7,5.2);
\node[flowlab] at (8.85,5.05) {read-only artifacts: typed, versioned, with lineage};
\draw[bi] (8.7,7.5) -- (8.7,7.8);
\node[flowlab] at (8.85,7.65) {tool calls and candidates; never conclusions};
\end{tikzpicture}}
\caption{Context as infrastructure. Knowledge compilation is the supply side: an ingestion pipeline that enriches raw documents into typed context artifacts. The harness is the control side: it decides which artifact, if any, enters the window and when the loop stops. Tools are the contract between them. Arrows show the flow of context, not control. The navigation tree (highlighted), whose leaves are claim--evidence pairs, is the mechanism measured in this paper; the ontology artifact (highlighted) is the one whose value is primarily commercial rather than benchmark-visible.}
\Description{A four-band architecture: retrieval harness at top, tools below, knowledge compilation below that, and raw corpus at the bottom, with upward arrows for ingestion, read-only artifacts, and tool calls; the claim-leaf navigation tree and ontology templates are highlighted.}
\label{fig:architecture}
\end{figure*}

This paper argues that the layer missing from this picture is \emph{knowledge compilation}: a configurable \emph{ingestion} pipeline that uses LLMs to enrich source documents with context that is not present in any chunk on its own---navigation trees whose leaves are atomic claims with verbatim evidence, entity relations, timelines, mind maps, ontology labels, synthesized articles, and durable memory. Compilation happens once, upstream of queries, and produces typed artifacts rather than a single index; retrieval then consumes those artifacts instead of raw text. Compilation is the supply side, the harness is the control side, and the tools are the interface between them (Figure~\ref{fig:architecture}).

The marginal value has moved from finding to supplying, and the centrality of compilation does not depend on beating search at recall---it does not. Under a competent harness we find that structural artifacts---RAPTOR-style summary trees, PageIndex-style outlines, LLM wikis, and entity--relation graphs---produce no detectable accuracy gain on FRAMES, and that a PageIndex-style table-of-contents index can become a negative asset at FRAMES scale. Structure is not additional evidence; what compilation provides is \emph{better context}, smaller, provenance-bound, and sufficient for the question at hand.

Two consequences organize the rest of the paper. First, compilation is the subject and the harness is the discipline that makes it usable: we define knowledge compilation as the ingestion-time supply of context, give a taxonomy of artifacts with the ledger on which each should be evaluated, and report a negative result with a mechanism---structural compilation does not improve accuracy under a competent harness, and a PageIndex outline becomes a liability beyond a scale we identify. Second, the mechanism we can measure is claim--evidence compilation: we specify it, give three serving modes, and derive a cost model with a 2\% break-even sufficiency rate, a 43\% implied FRAMES rate, an amortization rule, and the latency price of substitution.

\paragraph{Scope.} Our quantitative claims concern English text corpora, BM25/grep harnesses, and question answering. The deployment claims about ontologies, graphs, wikis, timelines, mind maps, and memory are reported on a separate ledger with named metrics and instrumentation (Table~\ref{tab:deployment}); they are not benchmark accuracies and we do not present them as controlled effects. Section~\ref{sec:limits} states the conditions under which we expect the quantitative conclusions to reverse.

\section{Related Work}
\label{sec:related}

\subsection{From RAG to the Search Harness}

Classical RAG conditioned generation on passages from a sparse or dense retriever~\citep{lewis2020rag,karpukhin2020dpr,izacard2022contriever}; its limits on multi-hop questions motivated iterative retrieval~\citep{trivedi2022interleaving}, self-reflection~\citep{asai2024selfrag}, and adaptive control over when to retrieve~\citep{jeong2024adaptiverag}. The harness turn goes further: the agent owns search. ReAct interleaves reasoning and tool calls~\citep{yao2023react}; enterprise AgenticRAG decomposed retrieval into SEARCH, FIND, and OPEN and showed that BM25 improves inside the loop~\citep{vijayaraghavan2025agenticrag}; Direct Corpus Interaction reduced retrieval to terminal tools---grep, file reads, shell commands---and outperformed dense, sparse, and reranking baselines on BrowseComp-Plus and multi-hop QA~\citep{li2026beyond}. ``Is Grep All You Need?'' reached a similar conclusion~\citep{subramanian2025grep}, a preregistered comparison found exact search approaching an LLM-compiled wiki~\citep{agrawal2026vectorrag}, and tuned BM25 with a capable agent reaches 83.1\% on the same benchmark~\citep{hsu2026piserini}. SIRA enriches the index with LLM-generated evidence vocabulary so that iterative search collapses into one lexical call~\citep{yang2026sira}. We accept this literature's finding rather than dispute it: if lexical search inside a harness is sufficient for accuracy, then any additional compilation layer must justify itself on a different axis, and this paper asks which axis, and for which artifact. The position on the vector index follows from the same evidence: inside a search harness a dense leg is not a neutral addition, and in our runs it lowered accuracy, which is the direction these reports find. We therefore run no dense-retrieval leg anywhere in this paper.
\subsection{The Compilation Family}

Knowledge compilation, named after the LLM-wiki pattern~\citep{karpathy2026llmwiki}---an LLM compiling a corpus into a maintained, linked artifact that can be read instead of retrieved---is used here in a broader sense, defined in Section~\ref{sec:compilation}. The name also has a formal tradition, in which a logical knowledge base is compiled into a target language on which queries are tractable~\citep{darwiche2002knowledge}; that is not our usage. Four families dominate pre-compilation. The first is \textbf{navigation trees}: \textbf{summary trees} that recursively cluster and summarize chunks, as in RAPTOR~\citep{sarthi2024raptor}, and \textbf{tables of contents} that parse headings into an outline, as in PageIndex~\citep{pageindex2025}. \textbf{Wikis} rewrite the corpus into maintained articles~\citep{karpathy2026llmwiki,pinecone2025nexus}. \textbf{Knowledge graphs} extract entities and relations, for global summarization~\citep{edge2024graphrag} or for structured reasoning~\citep{yasunaga2021qagnn,wang2021kepler}.

A separate line reduces context at query time. LLMLingua compresses prompts token-by-token or phrase-by-phrase~\citep{jiang2023llmlingua}, and lost-in-the-middle studies show that models underuse material in long contexts~\citep{liu2024lost}. Compilation is corpus-level and query-independent: the compression is paid once, reused, and keeps a pointer to the exact source span. Evidence-centric QA has long paired answers with supporting spans, and citation evaluation requires both correctness and attribution~\citep{gao2023alce,min2023factscore}. We move that discipline from the answer to the index: every claim is stored with the verbatim evidence that supports it.

Prior compilation work asks: \emph{does the artifact retrieve better?} We ask: \emph{given that the harness retrieves, what should the model read, and where does the artifact create value?} The first question yields null or negative accuracy results for structure in our study; the second yields a large cost effect for evidence and a set of deployment effects for the other artifacts.

\section{Knowledge Compilation as an Ingestion Layer}
\label{sec:compilation}

\subsection{Definition}

We define \emph{knowledge compilation} as a query-independent transformation
\begin{equation}
\kappa:\; \mathcal{D}\;\longrightarrow\;\mathcal{A}=\{a_1,\ldots,a_m\},
\end{equation}
where $\mathcal{D}$ is a corpus and each artifact $a_j$ is a typed, versioned function of source documents that preserves a lineage link to the exact spans from which it was derived. The name follows the LLM-wiki usage of compiling a corpus into a maintained artifact~\citep{karpathy2026llmwiki}; our definition covers the whole family of semantic enrichment operations, of which the wiki is one template. Compilation may be deterministic (parsing, chunking), model-based (claims, summaries, entities, ontology labels), or both. In the Tree template the artifact is one navigation tree: internal nodes organize, and the leaves are claim--evidence pairs.

\subsection{The Pipeline}

\textbf{Chunking is a retrieval necessity, not a semantic goal.} Traditional RAG treats chunk quality as a primary determinant of retrieval quality; that is true for naive RAG, where one lookup must return an answer-bearing passage and the boundary decides what can be found. We open the pipeline with this point because it fixes the scope of everything that follows: compilation is a supply layer that enriches what the harness has already located, not a new retrieval geometry, so none of our results depend on a special chunking strategy, and the comparison in Section~\ref{sec:results} holds the chunker fixed on both arms. The chunk is a unit of recall, not a unit of meaning: the impossible tuning problem in which every corpus needs its own chunk size does not survive the harness.

The pipeline has six stages, each replaceable.
\begin{enumerate}[leftmargin=1.8em]
  \item \textbf{Parse.} Convert PDF, Office, HTML, Markdown, tables, and logs into a canonical representation that preserves headings, lists, tables, and page anchors.
  \item \textbf{Chunk.} Produce chunks at natural boundaries. The chunker is configurable; the baseline harness can see the same chunks, so compilation is not confounded with chunking, and no template requires a chunk size of its own.
  \item \textbf{Enrich.} A model reads each chunk in document context and emits one or more artifacts: atomic claims with evidence, summary labels, entities and relations, events and dates, ontology labels, or a synthesized article. The extractor here is \texttt{glm-4-flash}.
  \item \textbf{Validate.} Check evidence spans against the source, entity types against the ontology, dates against the document, and schema against the template. Invalid units are dropped, not repaired by the model.
  \item \textbf{Assemble.} Cluster claims into a carrier tree, resolve entities into a graph, order events into a timeline, or link articles into a wiki.
  \item \textbf{Register.} Store each artifact with its source version, offsets, extractor version, validation status, and access-control labels, which is what makes fallback, citation, audit, and incremental update possible.
\end{enumerate}
The stages form a \emph{template}. A Tree template emits claim--evidence pairs and clusters them into a navigation tree; a Graph template emits entities and relations; an Ontology template adds classes, properties, and constraints; a Wiki template synthesizes articles; a Timeline template orders extracted events; a Mind-map template lays out a document's entities as a navigation graph; a Memory template extracts durable facts from session logs. Templates share the parser, validator, and registry, so a corpus can carry several artifacts at once and a deployment can switch between them without re-ingesting: a legal collection may compile a claim-leaf tree for citation and an ontology for privilege rules, a support collection a claim-leaf tree plus session memory.

\subsection{Artifacts and the Ledger on Which They Are Judged}

Table~\ref{tab:artifacts} lists the artifacts we have implemented, what each adds at ingestion, how retrieval uses it, and what our FRAMES study shows. The last two columns are deliberately separate: a benchmark measures accuracy and token cost on a fixed question distribution, while a deployment measures whether a domain expert can find, trust, and govern an answer. The two rows whose claim leaves are removed need a warning: they are carriers without payload, a reader can browse them and a maintainer can scope a change with them, but they contribute no evidence of their own, so their deployment value is organizational rather than evidential and must not be read as an independent reason to ship them.

\begin{table*}[t]
\centering
\caption{Compilation artifacts, the context they add at ingestion, and the ledger on which each is evaluated. FRAMES results are from the controlled harness in Section~\ref{sec:results}; deployment value is assessed with the metrics and instrumentation in Table~\ref{tab:deployment}, not with benchmark accuracy. The two carrier-only rows remove the claim--evidence leaves, so their deployment value is organizational only---they arrange content, they do not add evidence.}
\label{tab:artifacts}
\footnotesize
\begin{tabular}{>{\raggedright\arraybackslash}p{2.5cm}>{\raggedright\arraybackslash}p{4.3cm}>{\raggedright\arraybackslash}p{3.0cm}>{\raggedright\arraybackslash}p{1.9cm}>{\raggedright\arraybackslash}p{3.3cm}}
\toprule
\textbf{Artifact} & \textbf{Added at ingestion} & \textbf{Query-time use} & \textbf{FRAMES result} & \textbf{Deployment value} \\
\midrule
Navigation tree (claim--evidence leaves) & Atomic claims with verbatim spans as leaves; cluster labels above them & Prefetch leaves; substitution; drill down & acc.\ neutral; tokens $-38\%$ & Token cost; audit; organization \\
Navigation tree, RAPTOR summary nodes (claims removed) & Recursive summary nodes; claim--evidence leaves removed & Orientation; coarse-to-fine browsing & acc.\ neutral & Corpus overview; incremental maintenance; \emph{organization only, no evidence value} \\
Navigation tree, PageIndex outline nodes (claims removed) & Heading hierarchy; claim--evidence leaves removed & Route to a document or section; drill down & acc.\ neutral to negative & Human navigation; document QA; \emph{organization only, no evidence value} \\
Entity--relation graph & Entities, aliases, relations; no communities & PageRank chunks; fetch claim--evidence & acc.\ neutral & Related-source discovery; compliance \\
Ontology & Classes, properties, constraints, mappings & Typed retrieval; query expansion; validation & not measured & Vocabulary control; integration; governance \\
LLM wiki & Synthesized, linked articles & Background reading; synthesis & acc.\ neutral & Onboarding; documentation \\
Timeline & Events and dates organized for context & Temporal orientation & not measured & Audit trails; briefing; navigation \\
Mind map & Document entities laid out as a navigation graph & Concept orientation; browsing & not measured & Briefing; navigation \\
Session memory & Durable facts extracted from session logs & Personal and session recall & not measured & Continuity; support \\
\bottomrule
\end{tabular}
\end{table*}

\subsection{Context Quality and Channel Separation}

The value of compilation is twofold: it lowers the number of tokens needed to convey a fact, and it raises the quality of the context that carries it.

The last dimension depends on a strict rule: \textbf{raw and derived channels are separate}. Derived artifacts are read-only, typed, and never written back into the source, so a model reading a claim can fall back to the original text and an auditor can see that the claim is a derived object, not the document. This prevents the most common failure of LLM-maintained corpora---a generated summary becoming the ``source'' for the next generation---and it is a precondition for using compilation in regulated settings.

\subsection{What Compilation Is Not}

Compilation is not a replacement for search: it does not choose the query, rank candidates, or decide when the loop stops, and it is not a claim that structure improves recall---our own measurements say otherwise. And it is not free, which is why Section~\ref{sec:economics} derives when it pays.

\section{Claim--Evidence Pairs: A Measured Mechanism}
\label{sec:evidence}

We now specify the mechanism we can measure: a navigation tree whose leaves are claim--evidence pairs, built by claim--evidence compilation.

\subsection{Principle: Compile Evidence, Not Prose}

If compilation cannot add facts, it should not paraphrase them. The unit we compile is a \emph{claim--evidence pair}
\begin{equation}
u=\langle c,\;e,\;s,\;\tau\rangle,
\end{equation}
where $c$ is a normalized claim, $e$ is a verbatim evidence span that supports $c$, $s$ identifies the source chunk and character offsets, and $\tau$ records validity metadata such as document time. The evidence span is copied exactly from the source, and a claim without a locatable span is not stored, which separates the representation from a summary: every compiled unit can be checked, cited, and audited. The representation is lossy in prose and lossless in provenance, and for a chunk of a few hundred tokens the pairs typically occupy a small fraction of the original, which is the source of the savings in Section~\ref{sec:economics}.
\subsection{Construction}

\paragraph{Extract.} A small model reads each chunk and emits pairs under a template that asks for atomic, decontextualized claims and for the supporting sentence or clause copied verbatim. Extraction is per-chunk and embarrassingly parallel; in every experiment here the extractor is \texttt{glm-4-flash}, and we did not need a frontier model.

\paragraph{Validate.} Each evidence span is checked for exact presence in the source, and pairs whose span cannot be located are discarded. This is what makes a small extractor safe: the model may propose, but the source disposes. Validation also records offsets, so the pair can be highlighted in the original.

\paragraph{Assemble.} At ingestion, claims are embedded and clustered hierarchically. Internal nodes are navigational labels, not summaries; leaves are claim--evidence pairs. The tree is a carrier and an organizing device, not the retrieval mechanism. In the Tree template of Listing~\ref{lst:template}, clustering operates on claim embeddings and each leaf points back to its source chunk.

\paragraph{Cost.} Three properties keep compilation cheap: extraction is parallel and query-independent; the extractor can be small, because validation rather than model size protects quality; and artifacts are incremental, so when a document changes only its chunks are re-extracted, its claim set replaced, and affected clusters updated.

\begin{lstlisting}[caption={The Tree template used for all FRAMES experiments. Parser and chunker are the defaults seen by the uncompiled baseline; only extraction, validation, and assembly are compilation.},label={lst:template},float=t]
template: tree
parser:   default
chunker:  default
extract:
  unit: claim_evidence
  model: glm-4-flash
  require_verbatim_span: true
  validate: span_match
tree:
  leaves: evidence_units
  internal_nodes: embedding_clusters
  summary_nodes: false
serve:
  default: evidence_substitution
  fallback: raw_chunk
  gate: qwen-decision-model-preview
\end{lstlisting}

\subsection{Three Serving Modes}
\label{sec:serving}

\paragraph{E1: Evidence-first retrieval.} The harness searches the evidence index directly with lexical retrieval (BM25 over claim text, anchored on the chunks that lexical search already ranks) and receives evidence units as candidates. In our comparison, E1 outperformed locating a document and then descending its tree: the document was usually found correctly, the tree did not disambiguate within it, and the extra hop consumed a turn. Hierarchy is carrier, not retriever, and the retrieval is lexical throughout: the semantic step is the model's reformulation inside the loop.

\paragraph{E2: Evidence substitution (default).} The harness searches raw text exactly as it would without compilation. When it localizes a chunk, the projector replaces the chunk with that chunk's claim--evidence pairs before content enters the window, and a gate decides whether to answer from evidence or open the raw chunk. This preserves the retrieval behavior that already works and changes only reading cost; it is the default in RAGFlow.

\paragraph{E3: Decision-gated reading.} A lightweight decision model inspects the localized chunk and the question and decides whether a deep read is warranted. It is not a reranker: it controls reading depth, not candidate order. We use the Qwen decision model~\citep{qwen2026decisionmodel} (\texttt{qwen-decision-model-preview}). Without evidence, it saves about 30\% of tokens. The compiled artifact reduces what is read; the gate reduces when reading happens.
\subsection{The Economics of Substitution}
\label{sec:economics}

Let $T_0$ be the token cost of a query under the uncompiled harness, and let $p_h$ be the gate's \emph{acceptance rate}---the operational proxy for sufficiency. FRAMES gives the two conditional changes
\begin{equation}
\Delta_{\text{hit}}=-0.91,\qquad \Delta_{\text{miss}}=+0.019,
\end{equation}
relative to $T_0$. The expected fractional change is
\begin{equation}
\mathbb{E}[\Delta]=p_h(-0.91)+(1-p_h)(0.019)=0.019-0.929\,p_h.
\label{eq:expected}
\end{equation}
Compilation pays for itself in query tokens when $\mathbb{E}[\Delta]<0$:
\begin{equation}
p_h>p_h^\ast=\frac{0.019}{0.91+0.019}\approx2.0\%.
\label{eq:breakeven}
\end{equation}
This is the central economic fact of the mechanism. The measured FRAMES average of $-38\%$ corresponds, by Eq.~\eqref{eq:expected}, to
\begin{equation}
\hat p_h=\frac{0.38+0.019}{0.929}\approx43\%,
\end{equation}
which is plausible for a multi-hop benchmark in which many questions combine several chunks.

The cost structure is what keeps the penalty small: a miss costs one extra look at a chunk the harness has already found, while a hit replaces a full chunk read with a handful of evidence sentences.

\paragraph{Sensitivity of the break-even rate.} The break-even rate inherits the uncertainty of its two inputs, and it is far more sensitive to the miss penalty than to the hit saving, because the penalty is a small number. An absolute uncertainty of one and a half points in the penalty (0.4--3.4\%) moves $p_h^\ast$ between roughly 0.4\% and 3.6\%, whereas the same absolute uncertainty in the hit saving (89.5--92.5\%) moves it only between 2.0\% and 2.1\%. The 2\% figure is therefore a midpoint whose uncertainty is dominated by how often a projection fails, not by how much it saves when it works---and, for the same reason, the miss penalty is also the endpoint that a finite run pins down least precisely. That is a deployment instruction rather than a caveat: the quantity a team should measure carefully is the frequency of failed projections, not the size of the saving when a projection succeeds.

\paragraph{Amortization.} Let $C=c\,T_0$ be the compilation cost in query-equivalents. Over $N$ queries the net saving is $N\,\mathbb{E}[\Delta]\,T_0-c\,T_0$, and the break-even number of queries is
\begin{equation}
N^\ast=\frac{c}{0.929\,p_h-0.019}.
\label{eq:amortize}
\end{equation}
At $p_h\approx0.43$, $N^\ast\approx2.6\,c$: a corpus whose compilation costs ten query-equivalents pays back after about 26 queries. This is also why cheap extraction matters: reducing $c$ by an order of magnitude reduces the payback horizon by the same factor.

\paragraph{The no-compile boundary.} The 2\% threshold is an infinite-horizon statement, and it is routinely misread as a license to compile everything. For a finite deployment with $Q$ queries and compilation cost $c$ in query-equivalents, the net benefit is
\begin{equation}
B(Q,c,p_h)=Q\left(0.929\,p_h-0.019\right)-c,
\label{eq:netbenefit}
\end{equation}
so compilation pays only when
\begin{equation}
p_h>p_h^{\min}(Q,c)=\frac{0.019\,Q+c}{0.929\,Q}
=2.0\%+\frac{c}{0.929\,Q}.
\label{eq:nocompile}
\end{equation}
The second term is the whole decision: a pilot with $Q=1{,}000$ queries and $c=10$ query-equivalents needs $p_h>3.12\%$, not 2\%, and at a measured $p_h=2.5\%$ it still loses 5.8 query-equivalents---the asymptotic rule says compile, the finite rule says do not.

\begin{figure}[t]
\centering
\begin{tikzpicture}
\begin{axis}[
  width=\columnwidth, height=3.5cm,
  xbar, bar width=10pt,
  xmin=-125, xmax=30,
  xlabel={$\Delta$ total tokens (\%)},
  symbolic y coords={Hit,Miss,FRAMES avg.},
  ytick=data,
  axis lines*=left,
  xmajorgrids, grid style={black!10},
  tick label style={font=\scriptsize},
  label style={font=\scriptsize},
  enlarge y limits=0.22,
]
\addplot[fill=blue!55, draw=blue!65!black] coordinates {(-91,Hit)};
\addplot[fill=red!55, draw=red!65!black] coordinates {(1.9,Miss)};
\addplot[fill=black!35, draw=black!55] coordinates {(-38,FRAMES avg.)};
\node[font=\scriptsize, anchor=east] at (axis cs:-94,Hit) {$-91\%$};
\node[font=\scriptsize, anchor=west] at (axis cs:4,Miss) {$+1.9\%$};
\node[font=\scriptsize, anchor=east] at (axis cs:-41,FRAMES avg.) {$-38\%$};
\end{axis}
\end{tikzpicture}

\vspace{2mm}
\begin{tikzpicture}
\begin{axis}[
  width=\columnwidth, height=3.0cm,
  ybar, bar width=22pt,
  ymin=0, ymax=165,
  ylabel={Latency (s)},
  symbolic x coords={No gate,Decision gate},
  xtick=data,
  axis lines*=left,
  ymajorgrids, grid style={black!10},
  tick label style={font=\scriptsize},
  label style={font=\scriptsize},
  nodes near coords, nodes near coords style={font=\scriptsize},
  enlarge x limits=0.35,
]
\addplot[fill=black!35, draw=black!55] coordinates {(No gate,140)};
\addplot[fill=blue!55, draw=blue!65!black] coordinates {(Decision gate,30)};
\end{axis}
\end{tikzpicture}
\caption{Measured token and latency effects. Top: conditional change in total query tokens on FRAMES when projected evidence is sufficient (hit), insufficient (miss), and averaged over the benchmark. Bottom: end-to-end latency with and without the decision gate, with claim-leaf trees present. Negative bars are savings.}
\Description{Two bar charts: token change of minus 91 percent on hits, plus 1.9 percent on misses, and minus 38 percent on average; latency falls from 140 to 30 seconds with a decision gate.}
\label{fig:economics}
\end{figure}
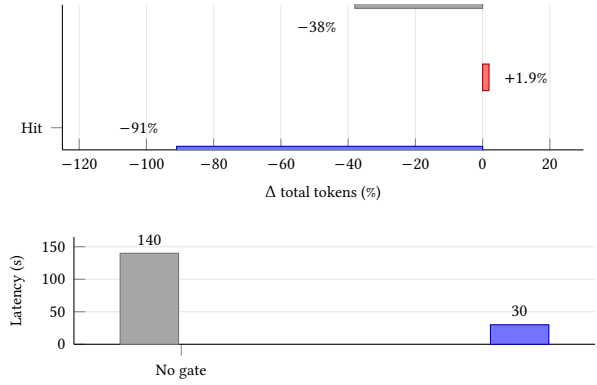

\section{What the Other Artifacts Buy}
\label{sec:artifacts}

The claim-leaf tree is the mechanism whose effect we can measure here. The other artifacts---navigation-tree carriers without claim leaves, graphs, ontologies, wikis, timelines, mind maps, memory---are not failed versions of it; they serve different consumers and different ledgers. This section states what each does well and why FRAMES does not see it.

\subsection{Navigation Trees: Organization, Not Recall}

A summary tree and a PageIndex outline are the same kind of artifact: a navigation tree. They differ in what the internal nodes contain---recursive summaries, document headings, or a human-oriented layout---and in how a reader moves through them, not in whether they navigate. RAGFlow changes what sits at the leaves: each leaf carries a claim--evidence pair, so reaching a leaf delivers evidence rather than another pointer. The structural baselines keep the carrier and remove that payload; the claim-leaf variant loads it, so the comparison is a leaf-content ablation, not a contest between different retrieval structures.

It does not add evidence by itself. Under a harness the agent can already localize with grep, so a static partition competes with the agent's next query and wins only if it is better; on heterogeneous text it usually is not. The right conclusion is not to delete the tree but to stop selling it as a retriever: a tree is a good place to \emph{put} evidence, it supports incremental maintenance, access control by subtree, and human navigation, and in our stack the carrier tree is exactly that---leaves are evidence, internal nodes are labels, and query-time search does the finding. Strip the claim leaves and the tree keeps the maintenance and the navigation and loses the evidence.

\subsection{Artifacts Without Evidence: Organizing What Extraction Already Found}

Three artifacts organize content that extraction has already produced, and none of them adds a fact. A \emph{timeline} orders events and dates so that the context given to the model has a chronology; its consumer is the context window, not the retriever. An \emph{LLM wiki} rewrites the corpus into linked articles and is good at synthesis, but it is bad at provenance unless every sentence links back, it goes stale unless recompiled, and on FRAMES it was neutral, so we treat it as a consumption layer over evidence. \emph{Memory} extracts durable state---preferences, decisions, commitments---from the session log rather than from documents, turning ephemeral interaction into queryable state~\citep{packer2023memgpt}; its evaluation is continuity rather than multi-hop QA.

A graph resolves entities and records relations that no chunk states in one place. It is deliberately not a GraphRAG-style community index, because community reports fragment knowledge and add a token-heavy layer of generated text. At query time a PageRank over the graph ranks chunks, and the harness then fetches those chunks' claim--evidence pairs first, so the graph improves recall of related material without paying to read raw text. FRAMES did not reward it, because the benchmark supplies entity-linked evidence and most questions are answerable from a few passages rather than a join.

An ontology is a schema: classes, properties, constraints, and mappings to source vocabularies~\citep{hogan2021knowledge,noy2001ontology}. At ingestion it does three jobs a flat extractor cannot: it \emph{types} entities and claims, so ``customer'', ``client'', and ``account'' become one class with aliases; it \emph{constrains} extraction, so a contract can have an obligation but a product cannot; and it \emph{links} artifacts across sources, so a clause and a ticket can be recognized as the same obligation. At query time it enables typed retrieval, relation-aware expansion, and validation of answers whose relation is not permitted. The value is commercial: consistent terminology across systems, cross-source integration, access control by type, domain adaptation without retraining, and an audit trail a compliance officer can read---structures that enterprise search has long used to bridge vocabulary and source systems~\citep{hawking2004enterprise,kruschwitz2017enterprise}.

\subsection{Two Ledgers, One Instrument}

Table~\ref{tab:deployment} makes the deployment ledger operational: each artifact is paired with one primary outcome, a metric with an explicit denominator, the place the metric is collected, and the rule that decides whether the artifact ships. The table is also a rejection criterion: an artifact whose metric does not move is reported as ``no measured effect''.

\begin{table*}[t]
\centering
\caption{The deployment ledger. Every row is defined for production collections, not FRAMES; the unit is the collection (or the reviewer team), and the comparison is against the same collection before the artifact or against a shadow control. Targets are pre-registered per deployment; a direction is reported only after the target is met, and audit coverage is an invariant rather than an average. The navigation-tree row refers to the claim-leaf variant; a carrier without claim leaves carries no evidence and is judged only on the organizational terms of Table~\ref{tab:artifacts}. The mind map is a navigation-tree layout and is judged on the navigation terms of the navigation-tree row, the timeline on the temporal terms of its own row.}
\label{tab:deployment}
\footnotesize
\begin{tabular}{>{\raggedright\arraybackslash}p{2.4cm}>{\raggedright\arraybackslash}p{3.3cm}>{\raggedright\arraybackslash}p{3.9cm}>{\raggedright\arraybackslash}p{3.2cm}>{\raggedright\arraybackslash}p{3.0cm}}
\toprule
\textbf{Artifact} & \textbf{Primary outcome} & \textbf{Metric (numerator / denominator)} & \textbf{Instrumentation} & \textbf{Decision rule} \\
\midrule
Navigation tree & Cheaper correct answer & tokens per correct answer; accepted answers per 1k queries & Harness token ledger tagged by station; judge verdict & Ship if tokens fall with non-inferior accuracy \\
Entity--relation graph & Related-source discovery & relational questions answered with $\ge2$ joined sources / relational questions; tokens per relational answer & Curated relational eval set; PageRank trace; token ledger & Ship if join success rises and tokens fall without audit loss \\
Ontology & Vocabulary control & entities mapped to a canonical class / extracted entities; unresolved-type rate & Compiler audit log; ontology validator & Ship if mapping target is met and invalid relations fall \\
LLM wiki & Onboarding and synthesis & time for a new user to answer 10 domain questions; citation follow-through rate & Task study; click-through to source spans & Ship only if time falls and citations stay traceable \\
Timeline & Temporal orientation & time to reconstruct event order; stale-version errors / temporal questions & Timeline validator; query logs & Ship if ordering accuracy rises and stale-version errors fall \\
Session memory & Continuity & repeat-question rate; corrected-fact rate; time to recover a prior decision & Session logs; user correction events & Ship if repeats and corrections fall \\
All artifacts & Auditability & delivered claims with a resolvable span and current source version / delivered claims & Citation binder; version register & Must be 100\%; otherwise block delivery \\
\midrule
\multicolumn{5}{@{}p{15.2cm}@{}}{\textbf{Status at writing.} No row of this ledger has yet met its pre-registered target in an instrumented deployment, so every row is reported as \emph{no measured effect} and none is promoted to a general claim; the auditability row is an invariant enforced by the delivery gate rather than a target we have met. We publish the ledger as a protocol with an empty result column rather than filling it with direction claims. A reported observation takes the form ``$C$ collections, $Q$ queries, metric from $A$ to $B$ ($D$\%), accuracy non-inferior, target met'', and is written only after the pre-registered target has been met on that collection set.} \\
\bottomrule
\end{tabular}
\end{table*}

Three measurement rules keep the ledger honest. First, the unit is the collection or the reviewer team, not the individual answer, because per-answer averages hide the collections where the artifact is unused. Second, the design is a shadow or A/B comparison on the same query mix, with the artifact enabled by collection hash, so traffic and question difficulty are not confounded with the treatment. Third, the target is written before the run, and a null is reported as a null.

The ledger in Table~\ref{tab:deployment} currently has no observations in it, and that absence is itself the report. A deployment claim in this paper means a named collection set, a pre-registered target, and an instrumented before/after or shadow comparison; we have no such observation, so the table states the measurement design and marks every row as \emph{no measured effect}. The shadow runs of Section~\ref{sec:implementation} produce exactly the labeled sample that a reportable observation requires; they are running, and no pre-registered target has been met yet.

\section{The Retrieval Harness: Discipline for the Loop}
\label{sec:harness}

\subsection{The Process Is Not the Answer}

An agentic loop can stop too early, answering from the first plausible passage, or never stop, accumulating evidence until the budget runs out. The second failure is more expensive: without discipline the loop rewrites the same query, rereads the same document, and fills the window with dozens of passages that are individually relevant and collectively useless. The process has consumed tokens, but it has not become an answer.

The harness exists to make the loop converge. Its job is not to make the model smarter or to give it every tool, but to guarantee properties a language model cannot guarantee for itself: that compiled evidence is fetched before the first call, that insufficient evidence is discarded rather than accumulated, that tools which do not work on this corpus are not offered, and that an abnormal stop still produces an answer. The framework owns the invariants---prefetch, sufficiency judgment, discarding an insufficient block, removing tools proven useless on this corpus, degrading model calls under budget, assigning citation identifiers at delivery, and composing a fallback answer on abnormal termination---because a model cannot be trusted to enforce its own budget or number its own citations reliably. The model owns the judgments that need semantics over underspecified language: choosing query terms and reading order, judging whether a passage is relevant, and stating what evidence is still missing. Mixing the classes---asking the model to enforce a budget, or the engine to judge relevance---is the usual source of unreliable agents.

\subsection{Tools by Failure Cause}

Tool lists are usually organized by implementation: search, browse, read, grep, graph. We organize tools by the failure they remove, with one primary failure per tool, one observable symptom, and one metric; if two tools remove the same failure, one is a configuration of the other, and if a tool removes no named failure, it is a demo. The failures we cover are vocabulary mismatch (BM25, grep, and claim-level evidence search), localization at scale (outline and tree navigation plus metadata scope), relations across sources (entity lookup, graph expansion, and chunk-keyed evidence lookup), temporal ambiguity (timelines, as-of filters, and a source-version register), synthesis burden (evidence projection and a wiki consumption layer), and tool or syntax error (a deterministic engine wrapper that turns errors into observations and disables the tool for the turn).

\paragraph{Navigation and graph tools are optional.} RAGFlow implements a navigation tool and a graph explorer. The navigation tool supports several strategies: aggregate documents from search results and refine to chunks through each document's own navigation tree (RAPTOR-style or PageIndex-style), or locate documents through claim and evidence search. In our benchmarks neither tool showed significant value. The weakest option we measured is the one now common in long-context RAG: let the harness read a navigation tree's table of contents and let the LLM choose which documents to read---very low accuracy at very high token cost---which surrenders the core value of a search harness, finding the answer quickly, accurately, and cheaply. Search will never die. Navigation tools are therefore optional in RAGFlow: on a corpus with no compiled artifacts the agent can use BM25 and grep alone, or also call the navigation tools, whose internals are a search-based aggregation; on a corpus with a compiled navigation tree they are not used at all, because claims and evidence projected onto the search results plus the sufficiency gate answer the same question at substantially lower token cost.

\paragraph{The zero-hit principle.} A zero-hit result is a property of the tool, not a fact about the corpus, and a harness that lets the model read ``0 results'' as absence has transferred an engine failure into a semantic conclusion. The harness explains the zero before the model sees it and retries along the fallback ladder---relax the filter, switch from compiled to raw lexical search, widen the time window---before the model may conclude that the corpus is silent. This small protocol prevents the most common false negative in agentic retrieval, where the agent gives up because a tool was misconfigured rather than because the answer is missing. A zero-hit explanation is a control message, not evidence, and it never receives a citation number.

\subsection{The Journey of a Query}

\paragraph{The five stations.} The query passes through five stations, sequential by default and skippable only by explicit configuration. \emph{(1) Capability check.} Before the model sees any tool, the harness asks whether this corpus has compiled artifacts at all: a corpus that was never compiled is not worth querying structurally, so navigation, graph exploration, and PageIndex outline tools are not exposed---not even their descriptions. \emph{(2) Evidence prefetch.} Before the model decides how to search, the framework performs one evidence recall across the whole corpus, as direct claim search or as raw search with substitution; either way it is a flat read, not a document route, because routing through a tree is biased---a small tree sends every question to the same one or two documents. \emph{(3) Evidence gate.} The harness checks sufficiency by count, source diversity, and coverage of the question; if the evidence suffices, one model call answers from it and the loop reads no raw text, and if it does not, the whole block is discarded rather than stuffed into the prompt. \emph{(4) Tool loop.} If the gate rejects the prefetch, the model enters the loop with the remaining tools; multiple retrieval actions in one turn execute in parallel, a step limit and a wall-clock budget make the loop converge by construction. \emph{(5) Deterministic close.} If the loop ends abnormally, the framework composes an answer from the question and the evidence already in hand.
\subsection{What the Framework Guarantees}

The stations imply a small set of invariants. (i) \emph{Evidence before thought}: compiled evidence is fetched before the first model call. (ii) \emph{Discard on insufficiency}: a rejected block is removed entirely, including from the system prompt and the citation table. (iii) \emph{Tool gating}: tools are exposed only if the corpus supports them and the tool has been useful there. (iv) \emph{Budgeted calls}: model calls are retried or degraded under a token and time budget. (v) \emph{Citations at delivery}: identifiers are assigned when evidence is delivered, not generated by the model. (vi) \emph{Guaranteed close}: every query ends with an answer or an explicit failure. These invariants are unglamorous, and that is the point: they are the difference between a demo that sometimes works and a system that can be operated.
\subsection{Not Free Tool Choice}

The design is deliberately opposed to the ``let the model choose among many tools'' style. Tool choice is a decision the model makes poorly when the toolset is large and the corpus unfamiliar; it consumes turns and produces confident but unproductive plans. The harness instead decides which tools are admissible, and the model decides how to use the admissible ones.

\section{Implementation}
\label{sec:implementation}

RAGFlow implements the design as a configurable ingestion pipeline feeding a stateful harness. The pipeline is a sequence of replaceable skills---parse, chunk, enrich, validate, assemble, register---and a template selects which artifacts are produced (Listing~\ref{lst:template}); the other templates reuse the same skills and write different artifacts.

\paragraph{Supply layer and harness.} Each compiled unit is stored with claim text, verbatim span, source chunk identifier and offsets, validity metadata, extractor version, and access labels; the store is versioned and content-addressed, so a claim resolves to the exact document version that produced it and can be highlighted in the source. The harness surfaces compiled units at the point of retrieval: claims are attached to the ordinary locate results, and a chunk-anchored lookup projects a recalled chunk to its claims. Navigation and graph tools are exposed only when the configuration calls for them: on a corpus with no compiled artifacts they are available as an optional aid, and on a corpus with a compiled navigation tree they are not used, because claims and evidence projected onto the search results answer the same question at lower token cost (Section~\ref{sec:harness}).

\paragraph{Claim grouping and judge selection.} Prefetched claims are grouped in document/page order and scored by a decision judge before admission. Of two decision models we evaluated---the Qwen decision model~\citep{qwen2026decisionmodel} and the StartLux decision model~\citep{startlux2026decision}---the Qwen-based one was far more reliable and faster (no failed groups in our runs, 30.5\,s total) than StartLux (18 failed groups, 560.1\,s), so it is the default. Both are small models released for exactly this role---deciding whether a block of evidence suffices---rather than general-purpose answering models. The harness therefore relies on row-level decision filtering and does not truncate pages by count alone (Section~\ref{sec:results}); when judge calls are slow or fail, it degrades gracefully.
\paragraph{Supply-layer seam and gate insertions.} The compiler and the harness meet at a small set of serving paths rather than one abstract interface: an existence probe over \texttt{compile\_kwd}, a chunk-anchored claim lookup, and a per-tool serving policy with three modes---off, compiled-only, and compiled-first with raw text on demand. The evidence gate is inserted where context can enter or leave the model---at prefetch admission, where the block is scored by unit count, source diversity, and entity coverage and either admitted with citations or discarded; at tool-result admission, where results are deduplicated, validated, and projected to units; and at delivery, where every claim must resolve to a delivered unit and a live source version. The harness never asks the supply side whether evidence is sufficient; that remains the gate's decision, which keeps the compiler replaceable and the gate testable without a compiler.
\paragraph{Shadow comparison.} A gate threshold cannot be tuned on the queries it rejects, because their counterfactual answer is unobserved. We therefore run the gate in shadow on a hash-bucketed share of production traffic: it computes and logs the decision but does not enforce it, and the query proceeds through the uncompiled path. A threshold is promoted to enforcement only when the accepted branch is non-inferior in accuracy and clears the token or latency target; otherwise the gate stays a logger.

\section{Evaluation}
\label{sec:results}

\subsection{Setup}

\paragraph{Benchmark.} FRAMES~\citep{krishna2024frames} contains multi-hop questions whose answers require facts from multiple Wikipedia articles, with entity-linked evidence. It is a hard setting for sufficiency: many questions cannot be answered from one chunk, so the gate should often fall back. We use FRAMES with the official answer protocol and an LLM judge.

\paragraph{Systems.} Every condition runs on FRAMES with the same harness, extractor, and gate; what varies across rows is the artifact or the serving mode. The uncompiled baseline is the harness with BM25 and grep, and it already reaches 91\% on FRAMES; the answering model is held fixed across conditions. The retrieval stack is the same in every condition: BM25 and grep over chunks, plus a lexical claim-text leg over the compiled evidence. No condition exposes a dense-retrieval leg: adding one lowered accuracy in our runs, consistent with the reports of Section~\ref{sec:related}, because the semantic work is done by the model's reformulation inside the loop, and the measurements in Table~\ref{tab:main} are therefore lexical throughout. For the structural baselines we implemented two navigation trees---one RAPTOR-style with summary nodes, one PageIndex-style with outline nodes---an LLM wiki, and an entity--relation graph, each exposed to the same harness as a tool. The two navigation trees are exposed for the structural audit of Section~\ref{sec:results} only; in the production serving configuration they are not used (Section~\ref{sec:harness}).

\paragraph{Sample sizes.} The token and accuracy figures of Table~\ref{tab:main} are measured on $N=200$ FRAMES questions, with the hit/miss split defined per query by the gate. The serving ablations of Section~\ref{sec:results} rest on fewer questions, and the latency comparison rests on the 7 questions that both arms answered. The counts that bear on how each result should be read are given here, so that the reader can weigh a number against its sample.

\paragraph{Metrics.} We report answer accuracy; total tokens per query, counting search calls, tool output, evidence, raw reads, and generation; sufficiency rate $p_h$; and end-to-end latency. Token counts are tokenizer-dependent, so we report relative changes except where noted.

\subsection{Structural Compilation: No Accuracy Gain}

On FRAMES, no compiled variant produced an accuracy improvement large enough to be distinguished from run-to-run variation in a single run. The two navigation-tree baselines are the same carrier with the claim--evidence leaves removed, and the claim-leaf variant loads them: the summary tree was neutral, the PageIndex outline neutral at small scale and harmful at FRAMES scale. The wiki was neutral and expensive to build, and it discarded the verbatim provenance needed for audit and exact-wording questions. The graph was neutral, because FRAMES already supplies entity-linked evidence and most questions do not require a join.

\begin{table*}[t]
\centering
\caption{Accuracy audit of the structural variants on the full FRAMES benchmark. The two navigation-tree rows share the same carrier; the only change is whether claim--evidence leaves are present. ``Neutral'' means no improvement large enough to be distinguished from run-to-run variation in a single run, and it does not rule out a small effect below the resolution of this design; ``negative'' means a loss of that kind. We report the direction and the failure mode the audit supports, not point estimates: each variant was run once. The claim-leaf row is the measured mechanism, not a structural baseline.}
\label{tab:structural}
\footnotesize
\setlength{\tabcolsep}{4pt}
\begin{tabular}{>{\raggedright\arraybackslash}p{3.0cm}>{\raggedright\arraybackslash}p{2.6cm}>{\raggedright\arraybackslash}p{7.0cm}>{\raggedright\arraybackslash}p{3.6cm}}
\toprule
\textbf{Variant} & \textbf{Accuracy vs.\ uncompiled} & \textbf{Failure mode observed} & \textbf{What the comparison isolates} \\
\midrule
Navigation tree, RAPTOR summaries (claims removed) & Neutral & Agent used summaries to orient, then grepped anyway; summaries added no facts and consumed a turn & Value of abstractive clusters without claim leaves, same carrier \\
Navigation tree, PageIndex outline (claims removed) & Neutral at small scale; negative at FRAMES scale & Extracted hierarchy incomplete or misaligned; a wrong branch cost more than open search; confident routing concentrated queries on few documents & Value of a static PageIndex outline without claim leaves, same carrier \\
LLM wiki & Neutral & Fluent synthesis, but lossy provenance and exact wording; expensive to rebuild & Value of prose compilation without spans \\
Entity--relation graph & Neutral & FRAMES supplies entity-linked evidence; few questions required a join & Value of cross-source relations on a single-source benchmark \\
Claim-leaf tree (claims loaded) & Neutral accuracy; $-38\%$ tokens & Evidence projection preserved spans and reduced reading; misses cost 1.9\% & Value of loading claim--evidence into the same carrier \\
\bottomrule
\end{tabular}
\end{table*}

The table is an audit, not a leaderboard. Its purpose is to make the negative result cumulative: two navigation-tree carriers without claim leaves, a wiki, and a graph, under the same harness, with no detectable accuracy gain. If structure were adding recall, at least one variant should have moved the accuracy endpoint; none did.

Four mechanisms explain the pattern. \emph{The harness is already adaptive}: grep followed by reading is an index built on demand, and a static partition wins only if it is better than the agent's next query. \emph{Structure is not evidence}: summaries and PageIndex outlines re-describe; they add no facts. \emph{Extraction errors compound}: a PageIndex outline that is 95\% correct per routing decision is unreliable across thousands---with branching factor eight, a 95\%-correct level yields about 75\% end-to-end routing success at $10^5$ documents, and a confident wrong branch is worse than open search. \emph{Structure adds a selection burden}: the agent must decide when to navigate, which consumes turns and context.

\subsection{Token Economics}

Table~\ref{tab:main} and Figure~\ref{fig:economics} report the positive result: when the gate judges projected evidence sufficient, total query tokens fall by 91\%, and when it does not they rise by 1.9\%, because the harness pays for a projection that did not suffice and then reads the chunk anyway. Averaged over FRAMES, tokens fall by 38\%, which by Eq.~\eqref{eq:expected} implies a sufficiency rate near 43\% against a break-even rate of about 2\%. The miss penalty is small because the projection is a lookup, not a generation; the hit saving is large because the baseline reads whole chunks while the projection reads sentences; and the average is dominated by $p_h$, which is a deployment property, since a corpus of mostly single-chunk questions has a higher $p_h$ and saves more.

\begin{table*}[t]
\centering
\caption{Main measurements on FRAMES with the Tree template and \texttt{glm-4-flash} extraction. Token changes are relative to the uncompiled BM25/grep harness; negative is a saving, and dashes indicate quantities not measured separately. The first three rows are conditional splits of the same benchmark---hit and miss are defined by the gate's verdict---and the FRAMES average is over all evaluated questions. The last two rows are arm means, not conditional splits: each is the mean change of that arm against the uncompiled baseline over the queries in that arm (evidence present, evidence absent), so $-52\%$ is not comparable with the $-91\%$ hit figure or with the $-38\%$ average. The arm-mean query counts are smaller still and are not reported separately.}
\label{tab:main}
\small
\begin{tabular}{>{\raggedright\arraybackslash}p{5.0cm}>{\raggedright\arraybackslash}p{2.7cm}>{\raggedright\arraybackslash}p{2.7cm}>{\raggedright\arraybackslash}p{1.9cm}>{\raggedright\arraybackslash}p{3.7cm}}
\toprule
\textbf{Condition} & \textbf{$\Delta$ tokens} & \textbf{Avg.\ tokens} & \textbf{Latency} & \textbf{Accuracy vs.\ baseline} \\
\midrule
Uncompiled (BM25/grep) & 0\% & $324{,}479$ & $46.7$\,s & 91.0\% (baseline) \\
Evidence sufficient (hit) & $-91\%$ & -- & -- & no reliable change \\
Evidence insufficient (miss) & $+1.9\%$ & -- & -- & no reliable change \\
FRAMES average & $-38\%$ & -- & -- & no reliable gain \\
Decision gate, with evidence (arm mean) & $-52\%$ & $155{,}305$ & $140\to30$\,s & no reliable change \\
Decision gate, no evidence (arm mean) & $-30\%$ & -- & -- & no reliable change \\
BrowseComp-Plus, harness only & -- & $\sim 3.0\times 10^6$/q & -- & 85.0\% \\
\bottomrule
\end{tabular}
\end{table*}

\subsection{Serving Ablations}

The ablations below select the serving configuration reported in Listing~\ref{lst:template}. They are stated as directions rather than as estimates, because their purpose is to choose a configuration, not to characterize a population.

\paragraph{Evidence-first versus document-then-tree, and substitution versus raw reading.} Direct lexical search over evidence units outperformed locating a document and then descending its tree: the document step was usually correct, but the tree did not reliably choose the right chunk within it, and the extra hop consumed a turn, so retrieval over evidence text does the finding and the tree organizes.

\paragraph{Recall ablations.} \textbf{Recall comes from anchor depth and from the union of the two lexical legs, not from searching harder.} We ablated claim retrieval on FRAMES and report the direction of each change rather than hit counts. Increasing anchor depth from 10 to 20 chunks improved hits, with page size growing accordingly; 30 gave no further gain. The union of the chunk-anchored leg and the claim-text leg was necessary: either leg alone left hits on the table. The configuration we kept is topN=20, $k=100$, the union of both legs, and no count-based truncation. There is no vector-depth ablation to report, because the harness has no dense leg to widen.

\paragraph{Page composition ablations.} \textbf{Truncating a page by count hurts recall, and a larger page does not reduce the number of rounds.} Fixed page caps (16 and 32) reduced recall, because relevant claims often appeared late and were removed. Doubling the cap also made the end-to-end turn worse rather than better: the larger page did not reduce the number of retrieval rounds, it only made each round more expensive, so the turn cost more tokens and more time than the smaller page it replaced. Reordering the page by query--claim lexical score did not improve hit rate. These ablations motivated avoiding count-based page caps and relying on row-level decision filtering.

\paragraph{The latency cost of substitution.}
The token saving is not free. On the seven questions answered by both arms of the ablation (one question failed in the uncompiled arm), the claim page roughly halved query tokens while end-to-end latency roughly tripled, and the mechanism is turns, not prompt size: a page that carries a smaller slice of evidence needs more retrieval rounds to converge. The gate is not the cause---judge waits account for only a small share of the claim arm's wall time---and the effect is a property of the representation, not of the configuration. Evidence substitution therefore buys tokens with turns: the same property that makes a page small makes it incomplete.

\subsection{Decision-Gated Reading}

With the claim-leaf tree present, the decision gate reduced mean end-to-end latency from 140\,s to 30\,s, and the gate arm of Table~\ref{tab:main} shows $-52\%$ tokens against the uncompiled baseline. As the caption of Table~\ref{tab:main} explains, that figure is an arm mean rather than the hit/miss split of the first three rows, and the gate's marginal contribution over substitution alone is modest because evidence is already compact. Without evidence, the same gate reduced tokens by about 30\% by declining deep reads that did not change the answer. The mechanisms address different costs: evidence reduces how much is read; the gate reduces how often a deep read happens.

\subsection{Gate Acceptance Is Not Answer Correctness}
\label{sec:gate}

We use $p_h$ as an operational sufficiency rate, but the quantity the gate actually produces is an \emph{acceptance} rate: the fraction of queries for which it admits the prefetched block. Acceptance is observable without a label; sufficiency is a judgment about whether the evidence could support the answer. A gate that accepts 43\% of queries may be right most of the time or may be admitting a fluent but irrelevant block; the acceptance rate alone cannot tell the difference, and neither can the token saving.

Let $A$ be acceptance, $H$ true sufficiency, and $C$ a correct final answer. Define the false-accept rate $\phi=\Pr(\neg H\mid A)$, the false-reject rate $\rho=\Pr(H\mid\neg A)$, and the conditional accuracies $a_T=\Pr(C\mid A,H)$ and $a_F=\Pr(C\mid A,\neg H)$. The accuracy on the accepted branch is
\begin{equation}
a_A=(1-\phi)\,a_T+\phi\,a_F,
\label{eq:accepted}
\end{equation}
so the false-accept rate is identified by
\begin{equation}
\phi=\frac{a_T-a_A}{a_T-a_F}.
\label{eq:falseaccept}
\end{equation}
This is the first-order estimator we use before a labeled audit. It is not identifiable from $p_h$ and token savings: those are one equation in two unknowns, and the denominator is the leverage. If a false accept answers 35 percentage points worse than a true accept, a 1\,pp gap between true-accept and accepted-branch accuracy implies $\phi\le2.9\%$, and a 3\,pp gap implies $8.6\%$. A gate can therefore look precise on accuracy and still route a substantial share of queries through the wrong evidence.

The operational consequence is that $p_h$ and $\phi$ must be reported together. We label a stratified sample of accepted queries for $H$, estimate $\phi$ from Eq.~\eqref{eq:falseaccept}, and treat the delivery gate as a separate check: acceptance says the block was admitted, delivery says every claim in the answer resolves to a live span. Neither is a guarantee of truth. The first controls what enters the context; the second controls what leaves it.

\paragraph{The minimum audit.} The estimator is only as good as the labeled sample behind it, so we specify the smallest audit that would close the loop, before running it. The labeled sample identifies $\phi$ directly, as the share of labeled accepted queries marked insufficient, with a binomial interval on that share: at 100 labels and a true rate near 10\% the interval spans roughly 4--16\%, which separates a gate that admits one wrong block in three from one that admits one in thirty but not 10\% from 20\%; at 400 labels the same interval spans about 7--13\%. Until that audit is run, Eq.~\eqref{eq:falseaccept} remains uncalibrated, and we report no value of $\phi$.

\subsection{Scale Check}

BrowseComp-Plus is a scale check for the baseline, not a compilation experiment. The harness-only configuration reaches 85\% on BrowseComp-Plus with BM25, grep, and MiniMax M3, at roughly three million tokens per question, and 91\% on FRAMES without any compiled index. The two figures are not comparable: BrowseComp-Plus is a single-question benchmark answered at roughly three million tokens, FRAMES is multi-hop and answered at roughly 325k, and each is a saturation check for the harness in its own setting. BrowseComp-Plus is also small enough that its size, rather than retrieval cost, limits what a compile-once experiment can show, so we do not use it to validate the token savings. The high accuracy nevertheless reinforces the negative result: accuracy is available without compilation, so a compilation layer has little accuracy headroom.
\subsection{Threats to Validity}
\label{sec:limits}

Our measurements are descriptive aggregates from one benchmark and one compiler model. We report no confidence intervals and no repeated seeds, so the 91\%/1.9\% endpoints are measured operating points, not population estimates. The ablations of Section~\ref{sec:results} select a configuration rather than estimate a population quantity, which is why they are stated as directions and why the one trade-off magnitude they establish---substitution's token saving against its latency cost---is given in round terms. FRAMES is English and Wikipedia-based; enterprise corpora with tables, scans, and jargon may have lower extraction quality, raising the miss rate and reducing savings. The PageIndex degradation is an observation at FRAMES scale, not a general law, and the 75\% routing figure is an analytic model rather than a measured scale sweep. The gate section reports $p_h$ but not a measured $\phi$: Eq.~\eqref{eq:falseaccept} is an estimator, and the stratified audit of Section~\ref{sec:gate}---100 to 400 labeled accepted queries---is specified but not run at benchmark scale. Deployment metrics are defined for instrumented production measurement, not a controlled effect, and we report a direction only where the pre-registered target was met.

\section{Discussion}
\label{sec:discussion}

\paragraph{What the sample sizes can and cannot support.} With few questions per ablation, we cannot estimate an effect to within a few percentage points, and we report no confidence intervals for those comparisons. What such samples can support is the direction and rough magnitude of effects large enough to survive small-sample noise: a token reduction near half, or a latency increase near threefold, is not plausibly an artifact of sampling, whereas a five-percentage-point difference could be. The headline figures of Table~\ref{tab:main} rest on a much larger sample and are reported as operating points; the ablation results are reported as directions for exactly this reason. The same asymmetry reappears in the break-even rate, which is dominated by the uncertainty of the miss penalty rather than of the hit saving (Section~\ref{sec:economics}), so the endpoint that most needs precision is the frequency of failed projections.

\paragraph{When compilation pays.} The economics reduce to five variables: the sufficiency rate, the miss penalty, the compilation cost, the query horizon, and the latency budget. Eq.~\eqref{eq:amortize} makes the trade explicit: a corpus whose compilation costs ten query-equivalents pays back after about 26 queries at $p_h\approx0.43$, and Eq.~\eqref{eq:nocompile} guards the opposite error, since a corpus with a short query horizon or an expensive compiler must clear a much higher sufficiency rate than the asymptotic 2\%.

The fifth variable is a constraint rather than a cost, and it can reverse the decision. Evidence substitution buys tokens with turns: on the paired subset of Section~\ref{sec:results}, query tokens roughly halved while end-to-end latency roughly tripled, because a page that carries a smaller slice of evidence needs more retrieval rounds to converge. When latency is a hard constraint, those token savings may not be worth buying, and the deployment should spend its budget on rounds instead: evidence-first retrieval removes the extra hop that document-then-tree routing costs, and decision-gated reading removes deep reads rather than adding them.

\paragraph{Design rules.} (1) Compile evidence, not prose, and keep a verbatim pointer for every claim: a summary cannot be audited and adds no facts, whereas a claim with a span makes fallback, citation, and audit possible. (2) Prefetch evidence flat and corpus-wide, never through a small tree, and gate sufficiency against a lossless fallback. (3) Measure conditional costs, not averages, and measure latency with them: a 38\% average hides a 91\% hit and a 1.9\% miss. (4) Compile only what the deployment will consume; the finite-horizon boundary of Eq.~\eqref{eq:nocompile} decides whether the consumption is enough. (6) Do not add a dense leg by default: in our runs it lowered accuracy, because inside a harness the agent reformulates, so a second ranking is something the loop must reconcile rather than a gain.
\paragraph{What an agent-native database is.} The phrase we want the reader to keep is \emph{agent-native database}, but the substance behind it is data supply. What an agent needs in order to answer well is high-quality context, and knowledge compilation is how that context is produced: claims with verbatim evidence, entity relations, timelines, ontology labels---typed, versioned, and governed. A system whose job is to produce and hand over that context, rather than to rank and return chunks, is an agent-native database. It is not a relational database: there is no query language and no schema enforcement, and the harness rather than the store decides what a given question reads. We report one compiler and one answering model, so the qualitative conclusion---evidence helps cost, structure does not help recall---is what we expect to generalize, not the exact percentages. The harness invariants are implemented in one system and validated by deployment, not by a controlled user study.

\section{Conclusion}
\label{sec:conclusion}

The retrieval layer is being absorbed into the search harness, and with it the old objective of beating lexical search on recall. Under a competent harness, structural compilation does not do that: summary trees, PageIndex outlines, wikis, and graphs produced no detectable accuracy gain on FRAMES, and a PageIndex outline became a liability at scale. Claim--evidence pairs, installed as the leaves of a carrier tree, cut total query tokens by 91\% when the evidence suffices and add 1.9\% when it does not, for a 38\% average, without a measurable accuracy loss and with a compiler that runs on an inexpensive small model---but the saving is bought with retrieval rounds, and the price is latency: on the paired subset of Section~\ref{sec:results}, query tokens roughly halved while end-to-end latency roughly tripled. The asymmetry makes the token economics work at a sufficiency rate of about 2\%, and a decision gate cuts latency from 140\,s to 30\,s; where response time binds, the gate and evidence-first retrieval are the levers and tokens are not the objective.

The broader claim is architectural: knowledge compilation is not one technique but a configurable ingestion pipeline that adds context upstream---claims for cost, trees for organization, graphs for relations, ontologies for governance, wikis for synthesis, timelines and mind maps for orientation, memory for continuity---while the harness decides what enters the window and the engine is the exact executor. Context becomes infrastructure when it is compiled once, typed, versioned, and governed, and when each artifact is judged on the ledger it was built for.

\paragraph{An agent-native database.} The architectural implication is that the system is no longer only a RAG or retrieval engine. Once context is compiled upstream, an agent no longer queries a ranked list of chunks; it is supplied with high-quality, provenance-bearing context that is compiled once and reused across queries, artifacts, and sessions. That supply function is what we mean by an agent-native database: the compiled artifacts are the data, the harness decides what enters the window, and the tools are the only path from stored context to model. It is not a relational database and not a new retrieval paradigm; it is a different answer to the question of what a retrieval system owes the agent---not a ranked list of passages to read, but the smallest sufficient set of verified facts, supplied in advance and governed like data.
\bibliographystyle{ACM-Reference-Format}
\bibliography{references}

@inproceedings{karpukhin2020dpr,
  author    = {Vladimir Karpukhin and Barlas Oğuz and Sewon Min and Patrick Lewis and Ledell Wu and Sergey Edunov and Danqi Chen and Wen{-}tau Yih},
  title     = {Dense Passage Retrieval for Open-Domain Question Answering},
  booktitle = {Proceedings of EMNLP},
  year      = {2020},
}

@inproceedings{izacard2022contriever,
  author    = {Gautier Izacard and Mathilde Caron and Lucas Hosseini and Sebastian Riedel and Piotr Bojanowski and Armand Joulin and Edouard Grave},
  title     = {Unsupervised Dense Information Retrieval with Contrastive Learning},
  booktitle = {Transactions on Machine Learning Research},
  year      = {2022},
}

@inproceedings{lewis2020rag,
  author    = {Patrick Lewis and Ethan Perez and Aleksandra Piktus and Fabio Petroni and Vladimir Karpukhin and Naman Goyal and Heinrich K\"{u}ttler and Mike Lewis and Wen{-}tau Yih and Tim Rockt\"{a}schel and Sebastian Riedel and Douwe Kiela},
  title     = {Retrieval-Augmented Generation for Knowledge-Intensive {NLP} Tasks},
  booktitle = {Advances in Neural Information Processing Systems (NeurIPS)},
  year      = {2020},
}

@inproceedings{yao2023react,
  author    = {Shunyu Yao and Jeffrey Zhao and Dian Yu and Nan Du and Izhak Shafran and Karthik Narasimhan and Yuan Cao},
  title     = {{ReAct}: Synergizing Reasoning and Acting in Language Models},
  booktitle = {Proceedings of ICLR},
  year      = {2023},
}

@inproceedings{asai2024selfrag,
  author    = {Akari Asai and Zeqiu Wu and Yizhong Wang and Avirup Sil and Hannaneh Hajishirzi},
  title     = {Self-{RAG}: Learning to Retrieve, Generate, and Critique through Self-Reflection},
  booktitle = {Proceedings of ICLR},
  year      = {2024},
}

@inproceedings{trivedi2022interleaving,
  author    = {Harsh Trivedi and Niranjan Balasubramanian and Tushar Khot and Ashish Sabharwal},
  title     = {Interleaving Retrieval with Chain-of-Thought Reasoning for Knowledge-Intensive Multi-Step Questions},
  booktitle = {Proceedings of ACL},
  year      = {2023},
}

@inproceedings{yasunaga2021qagnn,
  author    = {Michihiro Yasunaga and Hongyu Ren and Antoine Bosselut and Percy Liang and Jure Leskovec},
  title     = {{QA-GNN}: Reasoning with Language Models and Knowledge Graphs for Question Answering},
  booktitle = {Proceedings of NAACL},
  year      = {2021},
}

@inproceedings{wang2021kepler,
  author    = {Xiaozhi Wang and Tianyu Gao and Zhaocheng Zhu and Zhengyan Zhang and Zhiyuan Liu and Juanzi Li and Jian Tang},
  title     = {{KEPLER}: A Unified Model for Knowledge Embedding and Pre-trained Language Representation},
  booktitle = {Transactions of the Association for Computational Linguistics},
  year      = {2021},
}

@article{sarthi2024raptor,
  author    = {P. Sarthi and S. Abdullah and A. Tuli and S. Khanna and A. Goldie},
  title     = {{RAPTOR}: Recursive Abstractive Processing for Tree-Organized Retrieval},
  journal   = {arXiv preprint arXiv:2401.18059},
  year      = {2024},
}

@article{edge2024graphrag,
  author    = {D. Edge and H. Trivedi and J. Larson},
  title     = {From Local to Global: A Graph {RAG} Approach to Query-Focused Summarization},
  journal   = {arXiv preprint arXiv:2404.16130},
  year      = {2024},
}

@article{min2023factscore,
  author    = {Sewon Min and Kalpesh Krishna and Xinxi Lyu and Mike Lewis and Wen{-}tau Yih and Pang Wei Koh and Mohit Iyyer and Luke Zettlemoyer and Hannaneh Hajishirzi},
  title     = {{FActScore}: Fine-Grained Atomic Evaluation of Factual Precision in Long-Form Text Generation},
  booktitle = {Proceedings of EMNLP},
  year      = {2023},
}

@article{vijayaraghavan2025agenticrag,
  author    = {S. Vijayaraghavan and P. Agarwal and R. Chen},
  title     = {{AgenticRAG}: Agentic Retrieval for Enterprise Knowledge Bases},
  journal   = {arXiv preprint arXiv:2605.05538},
  year      = {2025},
}

@article{subramanian2025grep,
  author    = {Sathya Subramanian et al.},
  title     = {Is Grep All You Need? How Agent Harnesses Reshape Agentic Search},
  journal   = {arXiv preprint arXiv:2605.15184},
  year      = {2025},
}

@article{agrawal2026vectorrag,
  author    = {S. Agrawal and X. Li and Y. Chen},
  title     = {Vector {RAG} vs {LLM}-Compiled Wiki: A Preregistered Comparison on a Small Multi-Domain Research},
  journal   = {arXiv preprint arXiv:2605.18490},
  year      = {2026},
}

@inproceedings{hawking2004enterprise,
  author    = {David Hawking},
  title     = {Challenges in Enterprise Search},
  booktitle = {Proceedings of the Australasian Database Conference},
  year      = {2004},
}

@article{kruschwitz2017enterprise,
  author    = {Udo Kruschwitz and Charlie Hull},
  title     = {Searching the Enterprise},
  journal   = {Foundations and Trends in Information Retrieval},
  volume    = {11},
  number    = {1},
  pages     = {1--142},
  year      = {2017},
}

@article{li2026beyond,
  author    = {Zhuofeng Li and Haoxiang Zhang and Cong Wei and Pan Lu and Ping Nie and Yi Lu and Yuyang Bai and Shangbin Feng and Hangxiao Zhu and Ming Zhong and Yuyu Zhang and Jianwen Xie and Yejin Choi and James Zou and Jiawei Han and Wenhu Chen and Jimmy Lin and Dongfu Jiang and Yu Zhang},
  title     = {Beyond Semantic Similarity: Rethinking Retrieval for Agentic Search via Direct Corpus Interaction},
  journal   = {arXiv preprint arXiv:2605.05242},
  year      = {2026},
}

@article{yang2026sira,
  author    = {Zeyu Yang and Qi Ma and Jason Chen and Anshumali Shrivastava},
  title     = {Superintelligent Retrieval Agent: The Next Frontier of Agentic Retrieval},
  journal   = {arXiv preprint arXiv:2605.06647},
  year      = {2026},
}

@article{hsu2026piserini,
  author    = {Tz-Huan Hsu and Jheng-Hong Yang and Jimmy Lin},
  title     = {Rethinking Agentic Search with {Pi-Serini}: Is Lexical Retrieval Sufficient?},
  journal   = {arXiv preprint arXiv:2605.10848},
  year      = {2026},
}

@misc{karpathy2026llmwiki,
  author    = {Andrej Karpathy},
  title     = {{LLM} Wiki},
  howpublished = {\url{https://x.com/karpathy/status/1780884652155764942}},
  year      = {2026},
}

@misc{pinecone2025nexus,
  author    = {{Pinecone}},
  title     = {Introducing Nexus Knowledge Engine},
  howpublished = {\url{https://www.pinecone.io/blog/introducing-nexus-knowledge-engine/}},
  year      = {2025},
}

@inproceedings{krishna2024frames,
  author    = {Satyapriya Krishna and Kalpesh Krishna and Anhad Mohananey and Steven Schwarcz and Adam Stambler and Shyam Upadhyay and Manaal Faruqui},
  title     = {{FRAMES}: Factuality, Retrieval, And Reasoning {ME}asurement Set},
  booktitle = {Advances in Neural Information Processing Systems, Datasets and Benchmarks Track},
  year      = {2024},
}

@misc{pageindex2025,
  author    = {{Vectify AI}},
  title     = {{PageIndex}: Document Index for Vectorless, Reasoning-based {RAG}},
  howpublished = {\url{https://pageindex.ai}},
  year      = {2025},
}

@article{liu2024lost,
  author    = {Nelson F. Liu and Kevin Lin and John Hewitt and Ashwin Paranjape and Michele Bevilacqua and Fabio Petroni and Percy Liang},
  title     = {Lost in the Middle: How Language Models Use Long Contexts},
  journal   = {Transactions of the Association for Computational Linguistics},
  volume    = {12},
  pages     = {157--173},
  year      = {2024},
}

@inproceedings{jiang2023llmlingua,
  author    = {Huiqiang Jiang and Qianhui Wu and Chin-Yew Lin and Yuqing Yang and Lili Qiu},
  title     = {{LLMLingua}: Compressing Prompts for Accelerated Inference of Large Language Models},
  booktitle = {Proceedings of EMNLP},
  year      = {2023},
}

@inproceedings{gao2023alce,
  author    = {Tianyu Gao and Howard Yen and Jiatong Yu and Danqi Chen},
  title     = {Enabling Large Language Models to Generate Text with Citations},
  booktitle = {Proceedings of EMNLP},
  year      = {2023},
}

@inproceedings{jeong2024adaptiverag,
  author    = {Soyeong Jeong and Jinheon Baek and Sukmin Cho and Sung Ju Hwang and Jong C. Park},
  title     = {Adaptive-{RAG}: Learning to Adapt Retrieval-Augmented Large Language Models through Question Complexity},
  booktitle = {Proceedings of NAACL},
  year      = {2024},
}

@article{packer2023memgpt,
  author    = {Charles Packer and Sarah Wooders and Kevin Lin and Vivian Fang and Shishir G. Patil and Ion Stoica and Joseph E. Gonzalez},
  title     = {{MemGPT}: Towards {LLMs} as Operating Systems},
  journal   = {arXiv preprint arXiv:2310.08560},
  year      = {2023},
}

@article{hogan2021knowledge,
  author    = {Aidan Hogan and Eva Blomqvist and Michael Cochez and Claudio d'Amato and Gerard de Melo and Claudio Gutierrez and Sabrina Kirrane and Jos{\'e} Emilio Labra Gayo and Roberto Navigli and Axel-Cyrille Ngonga Ngomo and Axel Polleres and Sabbir M. Rashid and Anisa Rula and Lukas Schmelzeisen and Juan Sequeda and Steffen Staab and Antoine Zimmermann},
  title     = {Knowledge Graphs},
  journal   = {ACM Computing Surveys},
  volume    = {54},
  number    = {4},
  pages     = {1--37},
  year      = {2021},
  doi       = {10.1145/3447772},
}

@techreport{noy2001ontology,
  author      = {Natalya F. Noy and Deborah L. McGuinness},
  title       = {Ontology Development 101: A Guide to Creating Your First Ontology},
  institution = {Stanford Knowledge Systems Laboratory},
  number      = {KSL-01-05},
  year        = {2001},
}

@article{darwiche2002knowledge,
  title = {A Knowledge Compilation Map},
  author = {Darwiche, Adnan and Marquis, Pierre},
  journal = {Journal of Artificial Intelligence Research},
  volume = {17},
  pages = {229--264},
  year = {2002},
  doi = {10.1613/jair.989}
}

@misc{qwen2026decisionmodel,
  author    = {{Qwen Team}},
  title     = {Decision Model Preview},
  howpublished = {\url{https://help.aliyun.com/en/model-studio/decision-model-preview}},
  year      = {2026},
  note      = {Accessed 2026},
}

@misc{startlux2026decision,
  author    = {{StartLux}},
  title     = {Decision Model},
  howpublished = {\url{https://decision.startlux.com/}},
  year      = {2026},
  note      = {Accessed 2026},
}

\end{document}